\documentclass[aip, jcp, reprint, showkeys]{revtex4-1}
\usepackage{hyperref}
\usepackage{setspace}
\usepackage{array, tabularx, calc}
\usepackage{xcolor, color, colortbl}
\usepackage{amsmath, amssymb}
\usepackage{graphicx}

\newcommand{\new}[1]{#1}

\usepackage{soul}

\newcommand{\ilm}{Univ Lyon, Universit{\'e} Claude Bernard Lyon 1, CNRS, Institut Lumi{\`e}re Mati{\`e}re, 69622, Villeurbanne, France}
\newcommand{\erl}{Institute of Glass and Ceramics, Department of Materials Science, University of Erlangen-N{\"u}rnberg, Martensstr. 5, 91058 Erlangen, Germany}

\begin{document}

\title{Assessment of elastic models in supercooled water: a molecular dynamics study with the TIP4P/2005f force field}

\author{Emmanuel Guillaud}
\affiliation{\ilm}
\affiliation{\erl}
\author{Laurent Joly}
\affiliation{\ilm}
\author{Dominique de Ligny}
\affiliation{\erl}
\author{Samy Merabia}
\affiliation{\ilm}
\email[Corresponding author: ]{samy.merabia@univ-lyon1.fr}

\date{\today}

\begin{abstract}
Glass formers exhibit a viscoelastic behavior: at the laboratory timescale, they behave like (glassy) solids at low temperatures, and like liquids at high temperatures. Based on this observation, elastic models relate the long time supercooled dynamics to short time elastic properties of the supercooled liquid. In the present work, we assess the validity of elastic models for the shear viscosity and the $\alpha$-relaxation time of supercooled water, using molecular dynamics simulations with the TIP4P/2005f force field over a wide range of temperatures. We show that elastic models provide a good description of \new{supercooled water dynamics. For the viscosity, two different regimes are observed and the crossover temperature is found to be close to the one where the Stokes-Einstein relation starts to be violated. Our simulations show that only shear properties are important to characterize the effective flow activation energy.} This study calls for experimental determination of the high frequency elastic properties of water at low temperatures. 
\end{abstract}

\maketitle

\section{Introduction}%
Many materials display a viscoelastic behavior: they are elastic -- \textit{i.e.} they behave like solids -- at short times, whereas they are viscous -- \textit{i.e.} they flow -- at longer times. The distinction between liquid and solid is only conditioned by the timescale considered, the temperature and the level of stress applied to the material: under normal conditions, the Earth's mantle is solid at our timescale but flows on geological timescales, whereas water is liquid at our timescale and responds elastically at the picosecond scale~\cite{oswald2009rheophysics}. Glass former liquids are well known examples of such materials: at the laboratory timescale, they are glassy -- \textit{i.e.} solids -- at low temperature, whereas they flow like viscous liquids above the glass transition temperature $T_\text{g}$. Water falls in the category of glass formers, in the sense that it can be supercooled down to the homogeneous nucleation temperature $T = 231 \ K$~\cite{debenedetti2003supercooled}.

\smallskip%
One of the most puzzling features of the glass transition is the huge slowing down of the dynamics as probed \textit{e.g.} by the $\alpha$-relaxation time or the shear viscosity. It has been proposed that this slowing down is strongly related to the short time elastic properties of the supercooled liquid~\cite{dyre1996local, dyre2006colloquium, puosi2012correlation}, the correlation being described by the so-called elastic models. In the case of supercooled water, its slow dynamics has been characterized either experimentally or through molecular simulations~\cite{de2016mode, dehaoui2015viscosity, smith2000self, amann2016colloquium, kumar2007breakdown, moron2016macro, torre2004structural, gallo1996slow, guillaud2016decoupling}. 
However, very few works addressed the connection between the slow water dynamics and short times elastic properties. Recently, Klameth and Vogel discussed a connection between the $\alpha$-relaxation time and the high frequency shear modulus on the basis of Schweizer's elastic model~\cite{klameth2015slow}. We can also mention the study by Ciamarra \textit{et al.}, which does not concern specifically water, but a model glass former displaying a density maximum, as observed in water~\cite{ciamarra2015elastic}.

\smallskip%
The common idea behind elastic models is the following~\cite{dyre1996local, dyre2006colloquium, puosi2012correlation}. Collective relaxation processes such as flow events are infrequent in the supercooled liquid and occur on long timescales, consisting in local molecular rearrangements characterized by an effective activation energy $E_\text{a}(T)$. Elastic models relate the activation energy to the elastic properties of the supercooled liquid calculated on short -- picosecond -- time scales. In the shoving model for instance, the activation energy is proportional to the work $W$ necessary to shove the surrounding of a given molecule, so that flow can occur. The work $W$ depends on the elastic properties of the supercooled liquid, as described by the shear modulus, denoted $G$, and the bulk modulus, denoted $K$. If the local structure around a rearranging region has a spherical symmetry, the rearrangement consists in a purely shear deformation~\cite{dyre2004landscape}. Consequently, the observables characterizing the slow dynamics, including the shear viscosity $\eta$ or the collective structural relaxation time $\tau_\alpha$ -- generally denoted $X$ in the following, should be expressed as:
\begin{equation}
\left\{\begin{aligned}
&X = X_0\, \exp{\left(\frac{E_{\text{a}}(T)}{k_B T}\right)},\\
&E_\text{a}(T) = \lambda\, G_p(T)\, a^3
\end{aligned}
\right.
\label{eqn:elastic}
\end{equation}
where $k_B$ denotes Boltzmann's constant, $T$ is the temperature, $a$ denotes a molecular length,  $G_p(T)$ is the high frequency plateau modulus and $\lambda$ is a dimensionless number which takes values close to one (see Ref. \citenum{dyre2006colloquium} for a review of the various models leading to this expression). Another popular version of elastic models emphasizes correlations between the slow dynamics and the mean square
displacement (Hall-Wolynes equation)~\cite{ciamarra2015elastic, ottochian2009universal}:
\begin{equation}
X = X_0\, \exp{\left(\frac{a^2}{2\, \langle u^2(T) \rangle}\right)}.
\label{eqn:hallwolynes}
\end{equation}
where $\langle u^2(T) \rangle$ is the vibrational mean square displacement. Elastic models, in their two versions, proved to provide a good description of various systems including metallic, polymer and colloid glasses~\cite{wang2012elastic, mattsson2009soft, puosi2012correlation, hecksher2015review}, but showed also some limitations, as in the case of oxide glasses~\cite{potuzak2013dynamics}.

\smallskip%
In this paper, we assess the validity of the two versions of elastic models to describe the slow dynamics of supercooled water using molecular dynamics simulations. We consider the promising TIP4P/2005f force field which has been shown to reproduce accurately the water shear viscosity over a broad range of temperatures, extending deep in the supercooled regime~\cite{guillaud2016decoupling}. We report a strong correlation between the observables characterizing the dynamics and the high frequency shear modulus or the mean square displacement for the temperatures considered in this study. Conversely, elastic models perform well at low temperatures, \new{but for the viscosity two regimes with different effective molecular lengths $a$ are observed, with a transition temperature} 
on the order of 260\,K. Interestingly, this temperature is close to the one where the Stokes-Einstein relation begins to be violated.

\section{Numerical methods}%
All calculations were performed using the LAMMPS package~\cite{plimpton1995fast}. Water molecules were modeled by the TIP4P/2005f force field~\cite{gonzalez2011flexible}, which recently proved to be very precise to describe the shear viscosity and the self-diffusion coefficient of supercooled water~\cite{guillaud2016decoupling}. Cubic simulation boxes with periodic boundary conditions and containing respectively 8$^3$, 10$^3$, 12$^3$, 14$^3$, 16$^3$, 18$^3$, 20$^3$ and 22$^3$ molecules were equilibrated in the isothermal-isobaric ensemble under 1\,atm pressure and various temperatures between 225 and 360\,K, with four replica of the system for each box size. For the lowest temperatures considered, we did not detect any sign of ice nucleation. The self-diffusion coefficient, the shear viscosity, the $\alpha$-relaxation time and the high frequency shear modulus were then computed for each replica in the microcanonical ensemble. In detail, the self-diffusion coefficient was calculated from the slope of the molecular mean-squared displacement, the shear viscosity was computed within the Green-Kubo formalism, and the $\alpha$-relaxation time was obtained from the coherent intermediate scattering function using a procedure detailed in Ref.~\citenum{guillaud2016decoupling}. Table \ref{table_simulation_times} summarizes the simulation times at different temperatures.

\begin{table*}
	\caption{\label{table_simulation_times} Relaxation time $\tau_{\alpha}$ and simulation times at different temperatures. Note that at each temperature, we considered eight simulation box sizes and four replicas of the system for each size.}
	\begin{tabular}{|r|c|c|c|c|c|c|c|c|c|c|}
		\hline
		$T$ (K)              & $225$  & $240$  & $255$  & $270$  & $285$  & $300$  & $315$   & $330$   & $345$   & $360$   \\
		\hline
		$\tau_{\alpha}$ (ps) & $31.8$ & $18.3$ & $6.27$ & $3.04$ & $2.08$ & $1.08$ & $0.748$ & $0.498$ & $0.413$ & $0.376$ \\
		\hline
		simulation time (ps) & $500$  & $340$  & $300$  & $300$  & $300$  & $300$  & $300$   & $300$   & $300$   & $300$   \\
		\hline
	\end{tabular}
\end{table*}
 
\begin{figure}[!htb]
	\centering
	\includegraphics[width = \columnwidth]{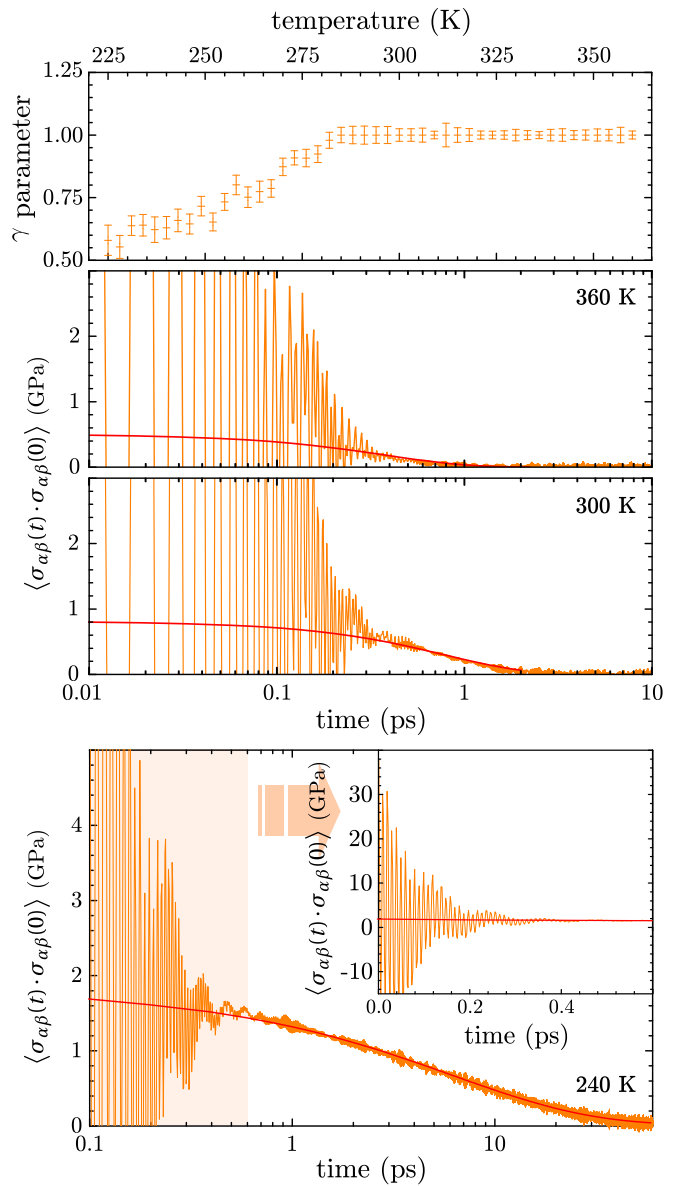}
	\caption{Extraction of the high frequency shear modulus from the shear stress auto-correlation function. A stretched exponential fit describes data at long timescales. Top: exponent $\gamma$ of the stretched exponential as a function of temperature. 
		Middle: Shear stress auto-correlation function at high temperatures, $T=360$\,K and $300$\,K. The solid lines show the stretched exponential fits.
		Bottom: Shear stress auto-correlation function at $T=240$\,K, together with the stretched exponential fit.
		Inset: the stretched exponential functional form is compared to the MD data at short timescales, emphasizing the relatively large amplitude of the oscillations of the stress auto-correlation function due to zero wave vector optical phonon modes~\cite{landry2008complex}.}
	\label{fig:graph01b}
\end{figure}%

Finally, we extracted the high frequency shear modulus from the autocorrelation function of the traceless stress tensor components $\langle\sigma_{\alpha\beta}(t)\, \sigma_{\alpha\beta}(0)\rangle$, where $\sigma_{\alpha\beta}$ denotes one of the five independent components of the traceless stress tensor ($\sigma_{xy}$, $\sigma_{yz}$, $\sigma_{xz}$, $(\sigma_{xx} - \sigma_{yy}) / 2$ and $(\sigma_{yy} - \sigma_{zz}) / 2$, see Ref.~\citenum{alfe1998first}). 
The plateau shear modulus $G_p$ is traditionally defined as the value of the elastic modulus where the derivative of $\log{(G(t))}$ is minimum~\cite{puosi2012correlation, dyre2012instantaneous}. The corresponding plateau should be located around 170\,fs, as inter-molecular connectivity bands appear around 200\,cm$^{-1}$ on the infrared absorption spectrum of bulk water at ambient temperature~\cite{brubach2005signatures}. 
However, due to zero wave vector optical phonon modes~\cite{landry2008complex}, the autocorrelation function of the stress tensor components strongly oscillates at short timescales (see Fig. \ref{fig:graph01b}), so that the plateau value $G_p$ must be extrapolated.
We fitted the stress autocorrelation function for $t > 0.03$\,ps using a stretched exponential functional form, %
\begin{equation}
\langle\sigma_{\alpha\beta}(t)\, \sigma_{\alpha\beta}(0)\rangle = \frac{G_0\, k_B T}{V}\, \exp{\left(-\left(\frac{t}{\tau_M}\right)^{\gamma}\right)} , 
\end{equation}
where $V$ denotes the volume of the system and $T$ the temperature, which was found to give good results over long time scales for each temperature. 
In the following, we will therefore use the parameter $G_0$ as a proxy to the high frequency plateau modulus $G_p$. Note that for the highest temperatures, extraction of $G_0$ becomes difficult as the oscillations of the stress tensor auto-correlation function extend over a time interval longer than the decay time $\tau_M$. The stretching exponent $\gamma$ as a function of temperature 
is shown in Fig.~\ref{fig:graph01b}. At high temperatures $T>280$\,K, the exponent $\gamma$ is close to $1$ and the stress autocorrelation 
function can be well described by a single exponential. Below $280$\,K, $\gamma$ decreases with cooling, indicating that a single Maxwell relaxation time 
can not describe stress relaxation at low temperatures, as also observed in Refs~\citenum{furukawa2011,kawasaki2017identification}.

Details about the different methods used are provided in Ref.~\citenum{guillaud2016decoupling}, where it is shown in particular that \new{for viscosity measurements,} the Green-Kubo formalism provides values in quantitative agreement with steady-state shear simulations.

\section{Results and discussion}%

\begin{figure}
	\centering
	\includegraphics[width = \columnwidth]{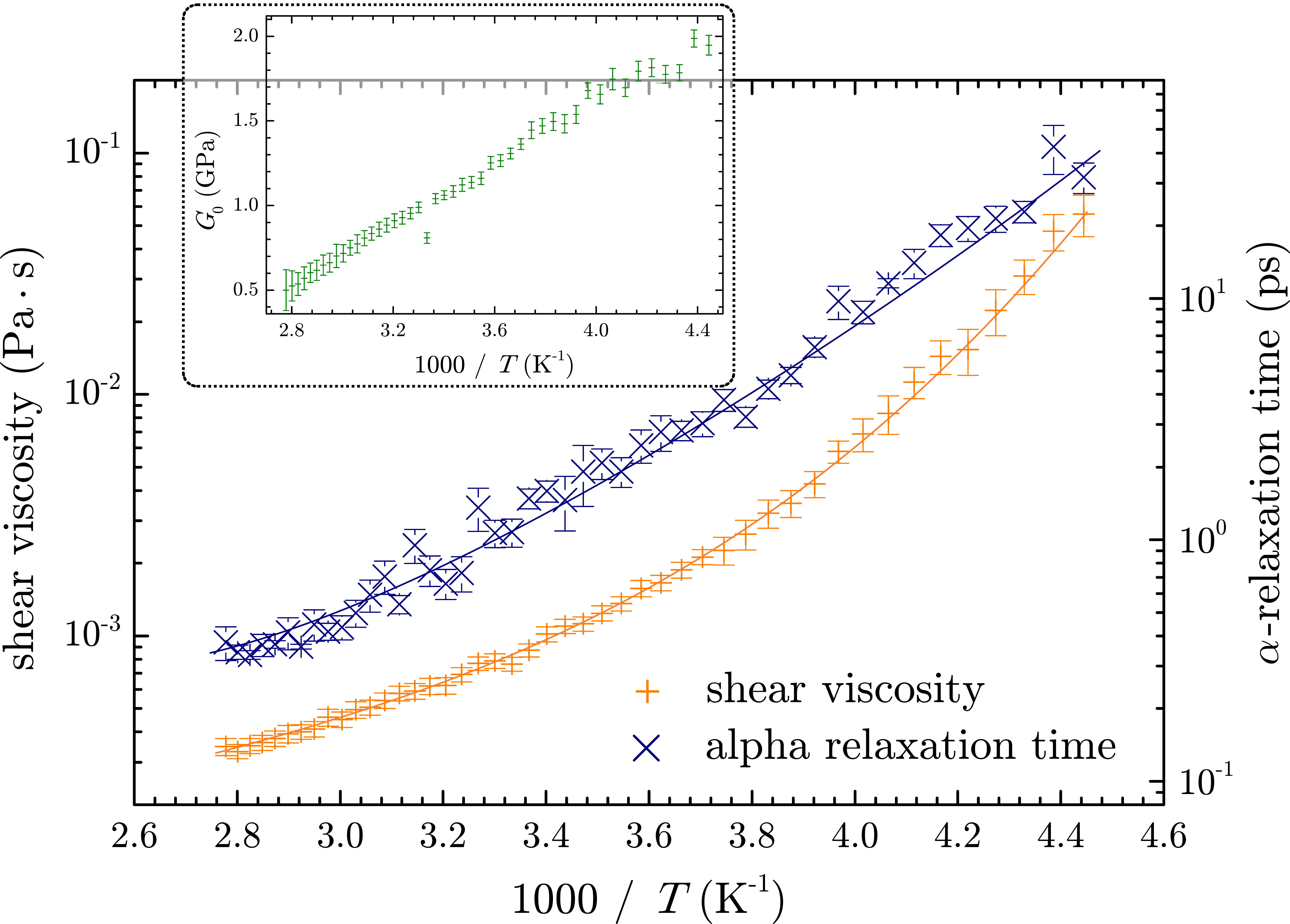}
	\caption{Evolution of the $\alpha$-relaxation time and the shear viscosity with respect to the temperature. Both observables display a non-Arrhenius behavior, although deviations to the Arrhenius law are smaller for the $\alpha$-relaxation time. Inset: temperature dependence of the shear modulus $G_0$.}
	\label{fig:graph03}
\end{figure}%
We first look at direct correlations between the shear viscosity or the $\alpha$-relaxation time of liquid water with the high frequency shear modulus across Eq.~\eqref{eqn:elastic}. %
As we showed in a recent paper~\cite{guillaud2016decoupling}, the TIP4P/2005f force field predicts a non-Arrhenius behavior for the shear viscosity and the $\alpha$-relaxation time, even if deviations to the Arrhenius law are smaller in the case of $\alpha$-relaxation time. Both observables are displayed in Fig.~\ref{fig:graph03}. The evolution of the high frequency shear modulus is also reported in the inset of this figure.
Interestingly, the shear modulus $G_0$ decreases monotonously with increasing temperature. 
Therefore, the temperature dependence of $G_0$ follows that of the plateau modulus $G_p$, and $G_0$ is clearly distinct from the infinite frequency shear modulus $G_{\infty}$~\cite{puosi2012correlation,dyre2012instantaneous}.

\begin{figure}
	\centering
	\includegraphics[width = \columnwidth]{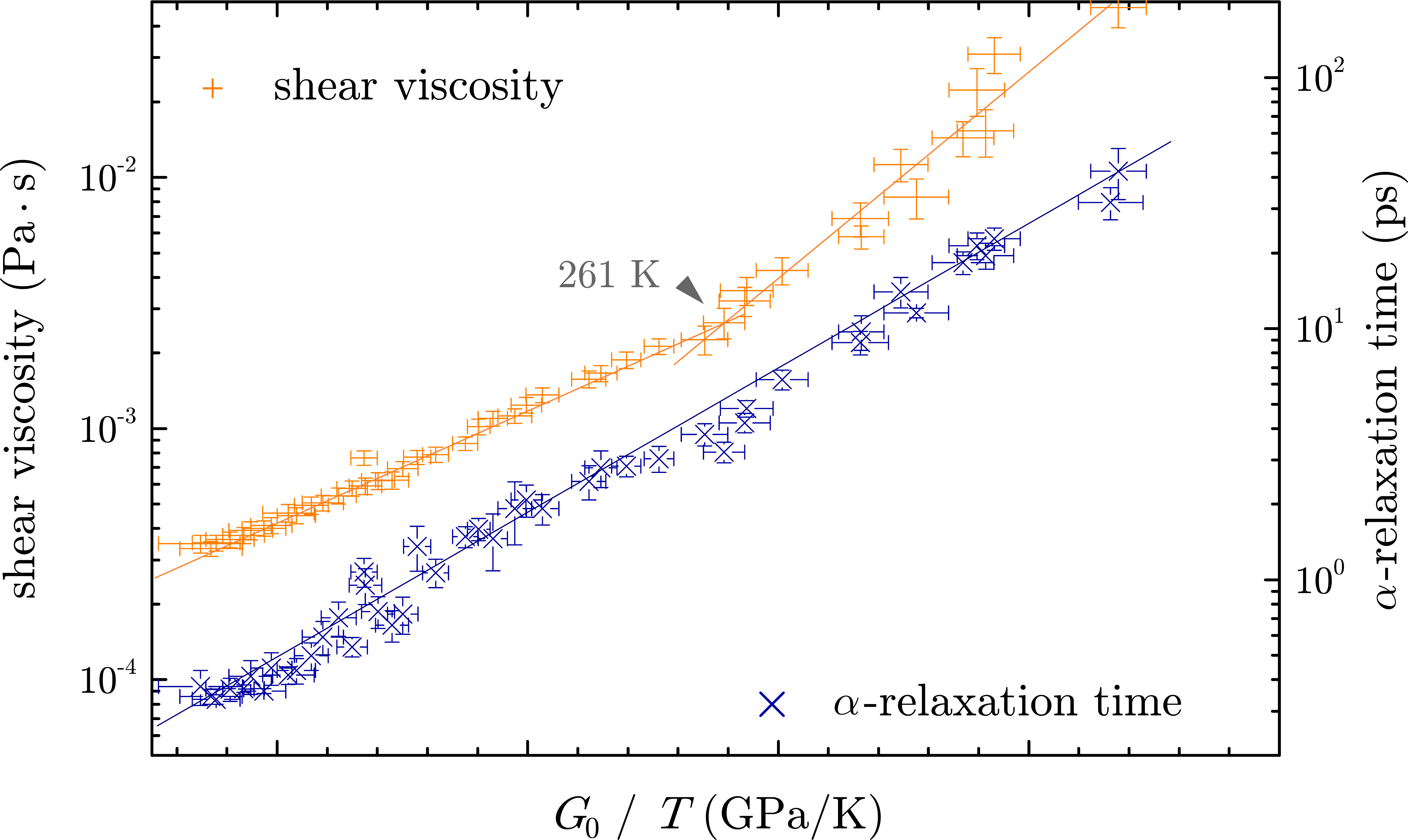}
	\caption{Correlations between the shear viscosity / $\alpha$-relaxation time and the high frequency shear modulus of liquid water. 
		While the $\alpha$-relaxation time can be described with Eq.~\eqref{eqn:elastic} over the whole temperature range, two different regimes are observed for the shear viscosity, with a transition temperature of ca. $261$\,K.}
	\label{fig:graph01a}
\end{figure}

Using the initial value $G_0$ of the stretched exponentials as a proxy of $G_p$, the agreement between elastic models and our calculations is only partial: for the shear viscosity, Eq.~\eqref{eqn:elastic} does not fit well the simulation data over the whole temperature range analyzed,
and two temperature domains can be distinguished, as shown in Fig. \ref{fig:graph01a}. In the two domains, Eq.~\eqref{eqn:elastic} describes the correlation between $G_0/T$ and the viscosity, but the parameter $a$ is not the same at high and low temperatures. For $T < 261$\,K, 
$a \approx 1.8\,\text{\AA}$, while for $T > 261$\,K, $a  \approx 1.5\,\text{\AA}$ (assuming $\lambda = 1$). 
\new{The fact that the parameter $a$ is larger at low temperatures may be explained by an increase of cooperativity, as discussed in the following paragraph.} 
%
As for $\alpha$-relaxation time, Eq.~\eqref{eqn:elastic} provides a good description of the slowing down of supercooled water with a single set of parameter $a \approx 1.6\,\text{\AA}$ over the entire temperature domain.

Interestingly, the temperature of transition between the high temperature and low temperature behavior of the viscosity is found to be close to 261\,K, which corresponds to the temperature below which the self-diffusion coefficient $D$ and the shear viscosity $\eta$ decouple with respect to the Stokes-Einstein relation, $D \eta \sim T$, which we estimated to be $265 \pm 15\,K$ and related to the appearance of dynamical heterogeneities in a previous work~\cite{guillaud2016decoupling}.
The Stokes-Einstein relation describes a regime where particle motion is not cooperative and may be compared to a Brownian particle diffusing in a bath of neighbor molecules. \new{Conversely, elastic models traditionally describe a situation where atomic motions involve collective displacements of the surrounding particles. However, here we saw that elastic models could still describe the slowing down of the dynamics at high temperatures, provided we defined an effective length $a$ which takes a smaller value as compared with the low temperature value. The fact that $a$ is smaller at high temperatures could then be related to the loss of cooperativity. Therefore, two regimes emerge in the description of the dynamics in the framework of elastic models: a diffusive homogeneous regime at high temperature where molecule displacements are almost independent and the Stokes-Einstein relation is obeyed, and a low temperature regime where molecule motions are cooperative.} %

\begin{figure}
	\centering
	\includegraphics[width = \columnwidth]{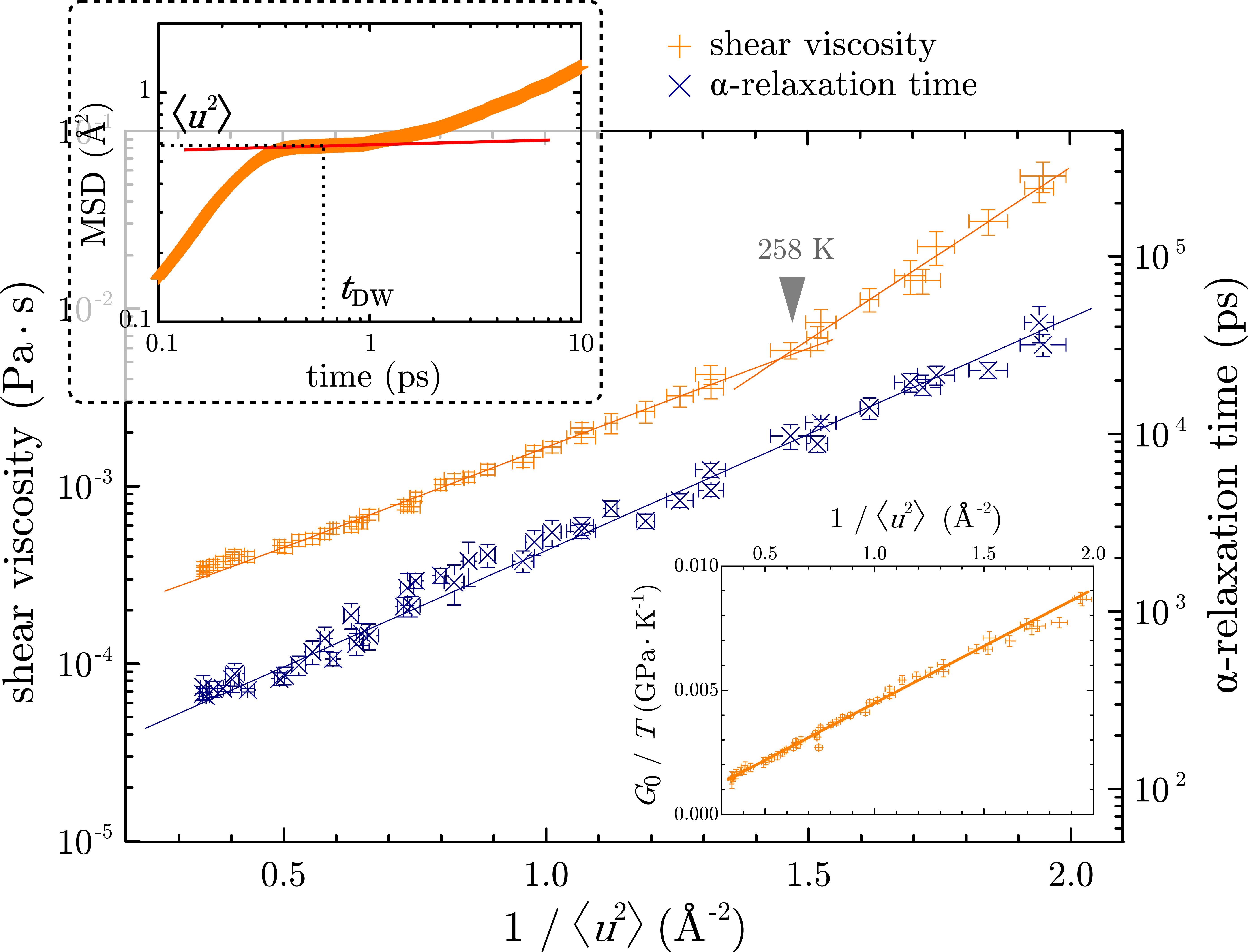}
	\caption{Correlations between the shear viscosity / $\alpha$-relaxation time and the vibrational mean squared displacement. The vibrational mean squared displacement is determined as the value of the mean squared displacement of the center of mass of a water molecule when the slope is minimum (see top left inset). Similarly to the previous correlations with the elastic modulus, the $\alpha$-relaxation time is described by Eq.~\eqref{eqn:hallwolynes} over the whole temperature range, while a transition is observed at \textit{ca.}~258\,K for the shear viscosity. The inset shows the correlation between the vibrational mean squared displacement and the high frequency shear modulus.}
	\label{fig:graph02}
\end{figure}

\smallskip%
We now assess the validity of the mean square displacement version of the elastic model, as described in Eq.~\ref{eqn:hallwolynes}. The quantity of interest here is the vibrational mean squared displacement, which can be accessed through the mean square displacement $\langle r^2(t) \rangle$, as illustrated in the top inset of Fig.~\ref{fig:graph02}. In a dense system, molecules move ballistically at short times while they diffuse at long times: $\langle r^2(t) \rangle = 6Dt$ where $D$ is the diffusion coefficient. At intermediate time scales, the mean square displacement of a molecule is limited by the presence of neighbouring molecules. As seen in fig.~\ref{fig:graph02}, at these intermediate time scales the MSD has a minimum slope, which corresponds to the time  $t_{DW}$ when the particle shoves its neighbors. The corresponding value of the MSD is called mean square vibrational amplitude or Debye-Waller (DW) factor~\cite{puosi2012correlation, dyre2004landscape}, and denoted $\langle u^2 \rangle$ in the following. %

Since the pioneering work of Buchenau and Zorn~\cite{buchenau1992relation}, Eq.~\eqref{eqn:hallwolynes} has been shown to hold in many systems~\cite{larini2008universal, puosi2012fast, ciamarra2015elastic}. For supercooled water, Fig.~\ref{fig:graph02} reports a good agreement between Eq.~\eqref{eqn:hallwolynes} and our simulation data at low temperatures. As for the previous elastic model, we find two different behaviors 
depending on the observable $X$. If $X$ is the $\alpha$-relaxation time, the agreement extends over the whole range of temperatures analyzed, with a molecular length $a = 1.62$\,\AA{}. In the case of the shear viscosity, 
\new{two regimes with different molecular lengths appear ($a = 1.97$\,\AA{} at low temperature, and $a = 1.48$\,\AA{} at high temperature), with a crossover temperature close to 258\,K.}   
This crossover temperature is very close to the one determined in the previous approach.
The equivalence between the two approaches discussed is confirmed by the inset of Fig.~\ref{fig:graph02}, which demonstrates the good correlation between $G_0/T$ and the MSD. 
The difference of behavior between the shear viscosity and the $\alpha$-relaxation time is not surprising, and can be related to the decoupling between these two observables observed at low temperatures~\cite{guillaud2016decoupling, shi2013se}.

The relevance of the vibrational MSD $\langle u^2 \rangle$ is supposed to hold while $\tau_\alpha$ is lower than $t_{DW}$~\cite{larini2008universal}, which corresponds to a temperature much higher than the crossover temperature $T=258$\,K. Actually, the activation energy associated with the two versions of the elastic models at this threshold temperature is close to the energy of a hydrogen bond (\textit{ca.}~20\,kJ/mol), which could explain the change of regime at high temperatures, since the connectivity of water molecules is mainly due to hydrogen bonds. At high temperature, thermal motion breaks the hydrogen bond network, so that the local tetragonal structure is short lived, and molecular displacements are no longer collective~\cite{kawasaki2017identification}. 

%
Before concluding, we should say a word about the role of quantum effects in the dynamics of supercooled water. Indeed, our simulations are fully classical and one may ask the relevance of quantum effects to describe the slow dynamics of water. These include the description of the nuclear degrees of motion, the hydrogen bond network but also the temperature dependence of the mean square displacement. We have already shown that the TIP4P/2005f potential reproduces accurately the shear viscosity of supercooled water~\cite{guillaud2016decoupling}. It would be interesting to measure the elastic properties of supercooled water at low temperatures, to see if the TIP4P/2005f potential can also describe the temperature dependence of the shear modulus, or the vibrational mean square displacement. This would allow us to conclude whether quantum effects do play a signicant role in the slow dynamics of water, or if they play a negligible role due to compensation effects. If quantum effects turn out to be important, demanding ab initio simulations could be unavoidable to investigate supercooled water dynamics\cite{ceriotti2016nuclear, gillan2016perspective, distasio2014individual}.%
\section{Conclusion}%
In conclusion, we show here that the long time dynamics of liquid water at low temperatures is highly correlated with its short time elastic properties, quantified by the shear modulus at high frequency. In particular, it is surprising to see that in spite of the numerous anomalies of water, elastic models perform as well as for simple glass formers including metallic glasses or supercooled polymers. Moreover, 
\new{for the viscosity two different temperature regimes with different effective molecular lengths $a$ are observed, and the crossover temperature is close to the one where the Stokes-Einstein relation starts to be violated.} 
These results call for experimental investigation of the high frequency shear modulus of supercooled water using \textit{e.g.}~inelastic X-ray scattering. Looking for correlations between the elastic modulus and the shear viscosity measured at low temperatures may also help in assessing the role of quantum effects which are not considered in our classical approach. \new{Finally, measuring the high frequency elastic properties of water may shed light on the molecular mechanisms relevant to interfacial thermal transport at solid-water interfaces~\cite{Merabia2016}.}

\begin{acknowledgments}
This study was performed using the computing resources of the PSMN, Computing Center of the Ecole Normale Sup{\'e}rieure in Lyon, and benefits from financial support of DFH/UFA~\cite{ufa}.
\end{acknowledgments}
\bibliography{references}%

\end{document}